\documentclass[numbers,sort&compress]{IntechOpen-Book}

\graphicspath{{Artworks/}}
\usepackage{url}
\usepackage{pgfplots}

\begin{document}

\Mainmatter

\begin{frontmatter}

\chapter{The Theory behind Controllable Expressive Speech Synthesis: a Cross-disciplinary Approach}
\author{No\'e Tits}
\author{Kevin El Haddad}
\author{Thierry Dutoit}

\makechaptertitle



\chaptermark{The Theory behind Controllable Expressive Speech Synthesis: a Cross-disciplinary Approach}

\begin{abstract} 

As part of the Human-Computer Interaction field, Expressive speech synthesis is a very rich domain as it requires knowledge in areas such as machine learning, signal processing, sociology, psychology.

In this Chapter, we will focus mostly on the technical side. From the recording of expressive speech to its modeling, the reader will have an overview of the main paradigms used in this field, through some of the most prominent systems and methods.

We explain how speech can be represented and encoded with audio features. We present a history of the main methods of Text-to-Speech synthesis: concatenative, parametric and statistical parametric speech synthesis. Finally, we focus on the last one, with the last techniques modeling Text-to-Speech synthesis as a sequence-to-sequence problem. This enables the use of Deep Learning blocks such as Convolutional and Recurrent Neural Networks as well as Attention Mechanism.

The last part of the Chapter intends to assemble the different aspects of the theory and summarize the concepts.

\end{abstract}

\begin{keywords} 
Deep Learning, Speech Synthesis, TTS, Expressive speech, Emotion
\end{keywords}


\end{frontmatter}

\section{Introduction} 

Controllable Expressive Speech Synthesis is the task of generating expressive speech from a text with control on prosodic features. 

This task is positioned in the emerging field of affective computing and more particularly at the intersection of three disciplines: 

\begin{itemize}

    \item Expressive speech analysis (Section 2), which provides mathematical tools to extract useful characteristics from speech depending on the task to perform. Speech is seen as a signal, such as images, text, videos or any kind of information coming from any source. As such, it can be characterized by a time series of features. 

    \item Expressive speech modeling (Section 3), modeling human emotions and their impact on the speech signal. Speech is considered here as a means of communication between humans.


    \item Expressive speech synthesis (Section 4), for which machine learning tools have become ubiquitous, especially hidden Markov models (HMMs) and more recently Deep Neural Networks (DNNs). 
    The field of Machine Learning allows machines to learn solving a given task. This field borrows from an ensemble of statistical models allowing to represent or transform data. It also uses concepts from Information Theory to measure distances between probability distributions.

\end{itemize}

\section{Expressive Speech Analysis}


\subsection{Digital Signal Processing}
\label{filter}

A signal is a variation of a physical quantity carrying information. The acoustic speech signal is converted into an electrical signal by a microphone. An acoustic signal is a variation of pressure in a fluid that the human perceives through the sense of hearing. This signal is mono-dimensional because it can be represented by a mathematical function with a single variable: pressure.\\

\begin{figure}[!ht]
	\centering
	\includegraphics[width=1\columnwidth]{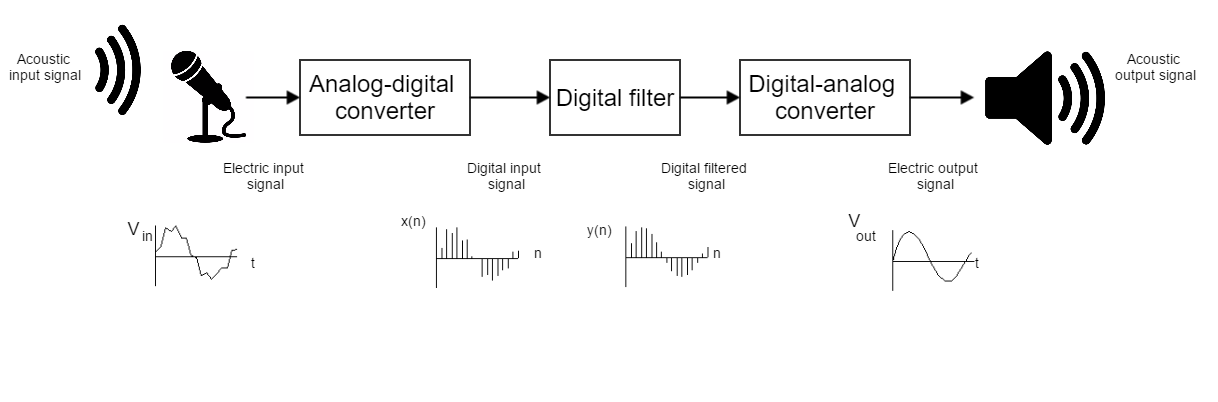}
	\centering
	\caption{Digital signal processing for acoustic signals}
	\label{filtering}
\end{figure}

The electrical signal generated by the microphone is an analog signal. In order to process it with a digital machine, it must be digitized. This is done by electronic systems called analog-to-digital converters that sample and quantify analog signals to convert them into digital signals. After some processing of the digitized signal, a digital-to-analog converter can be used to convert the processed digital signal back into an analog signal. This analog electrical signal can then be converted into an acoustic signal though loudspeakers or earphones to make it available to human ears. These steps are represented in Figure \ref{filtering}.\\

Digital signal processing~\cite{TcTs} is the set of theories and techniques for analyzing, synthesizing, quantifying, classifying, predicting, or recognizing signals, using digital systems.\\

A digital system receives as input a sequence of samples $ \{x (0), x (1), x (2), ...\} $, noted as $ x(n) $, and produces as output a sequence of samples $ y (n) $ after application of a series of algebraic operations.\\

A digital filter is a linear and invariant digital system. Let us consider a digital system that receives the sample sequences $ x_1 (n) $ and $ x_2 (n) $ as input. This system will respectively produce the sample sequences $ y_1 (n) $ and $ y_2 (n) $ as output. This system is linear if it produces the output $ \alpha y_1 (n) +\beta y_2 (n) $ when it receives the sequence $ \alpha x_1 (n) + \beta x_2 (n) $ as input.
A digital system is said to be invariant if shifting the  input sequence by $ n_0 $ samples also shifts the output sequence by $ n_0 $ samples. \\

These linear and invariant digital systems can be described by equations of the type:
\begin{equation}
 y(n)+a_1y(n-1)+a_2y(n-2)+...+a_Ny(n-N)=b_0x(n)+b_1x(n-1)+...+b_Mx(n-M) 
\end{equation}
or
\begin{equation}
y(n)+\sum_{i=1}^{N}a_iy(n-i)=\sum_{i=0}^{M}b_ix(n-i)
\label{filter_eq}
\end{equation}


This is equivalent to saying that the output $ y (n) $ is a linear combination of the last N outputs, the input $ x (n) $, and the M previous inputs. A digital filter is therefore determined if the coefficients $ a_i $ and $ b_i $ are known.  A filter is called non-recursive if only the inputs are used to compute $ y (n) $. If at least one of the previous output samples is used, it is called a recursive filter.

\subsection{Speech Features}

Speech is a signal carrying a lot of information. These expend from the sequence of words used to create a sentence, to the tone of voice used to utter this sentence. Not all of them are necessary to process and for some systems, trying to process all of them can harm the efficiency of the system. Also, the speech can carry noise before reception. That's why an important step in speech analysis, is to extract descriptors or features that are relevant to the task of interest.

There exist many different feature spaces that describe speech information. In this Section, we give an intuitive explanation of the ones widely used in Deep Learning architectures. 




\subsubsection{Power spectral density and spectrogram} \label{spectrogram}

 Fourier analysis demonstrates that any physical signal can be decomposed into a sum of sinusoids of different frequencies.

The power spectral density of a signal describes the amount of power carried by the different frequency bands of this signal.

This range of frequencies may be a discrete value set or a continuous frequency spectrum. In the field of digital signal processing, this power spectral density can be calculated by the Fast Fourier Transform (FFT) algorithm.\\

The graph of the power spectral density allows to visualize the frequency characteristics of a signal such as the fundamental frequency of a periodic signal and its harmonics. A periodic signal is a signal whose period is repeated indefinitely. The number of periods per unit of time that repeats is the fundamental frequency. Harmonics are the multiple frequencies of the fundamental. These frequencies have an important power density and present therefore extrema in the power spectral density.\\

\begin{figure}[!h]
	\centering
	\includegraphics[width=1\columnwidth]{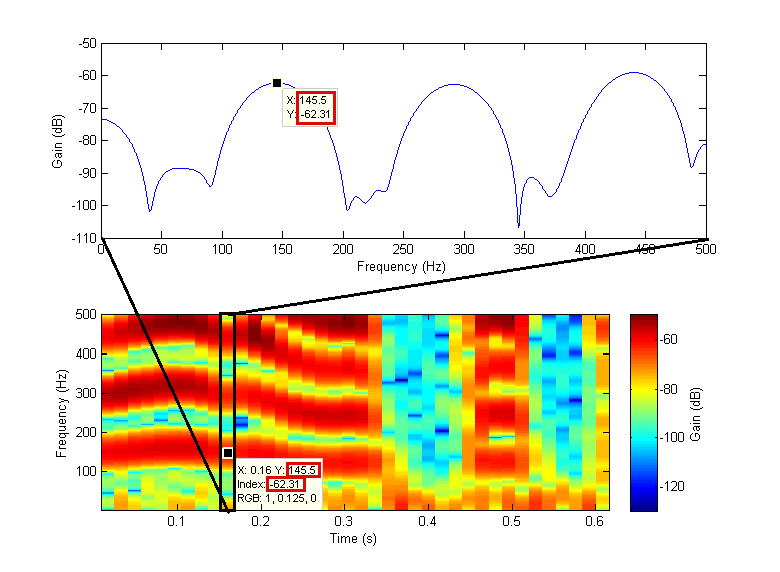}
	\centering
	\caption{Spectrum (top) and spectrogram (bottom) of a speech segment}
	\label{spectrum_spectrogram_example_edited}
\end{figure}

An example of power spectral density is shown in the upper part of Figure~\ref{spectrum_spectrogram_example_edited}. The first maximum is at the fundamental frequency which is 145.5 Hz. The others maxima are the harmonics.\\

When the signal's characteristics are evolving in the time, like with the voice for example, the spectrogram can be used to visualize this evolution. The spectrogram represents the power spectral density over the time. An example of power spectrogram is shown in the lower part of Figure~\ref{spectrum_spectrogram_example_edited}. The x-axis is time and the y-axis is frequency. The colors correspond to the power density. A color scale is given on the right of the graph. The spectrogram is thus constructed by juxtaposing power spectral density functions computed on every frame as suggested in Figure \ref{spectrum_spectrogram_example_edited}.\\

\subsubsection{Mel-Spectrogram}

The Mel-Spectrogram is a reduced version of the spectrogram. The use of this feature is very widespread for machine learning-based systems in general and for Deep learning-based TTS in particular.

The intuition behind this feature is to compress the representation of the speech in the higher values of the frequency domain based on the fact that human ear is sensitive to some frequencies more than others. The mel scale is an experimental function representing the sensitivity of human ear depending on the frequency.


The conversion of frequency $f$ in mel-frequency $m$ is:

\begin{equation}
     m=2595\cdot \log _{10}\left(1+{\frac {f}{700}}\right)
\end{equation}

Figure~\ref{mel_scale} shows the curve of the mel scale as a function of the frequency. As one can observe, an interval of low frequencies is mapped to a larger interval of mel values than for high frequencies. As an example, the interval $[0, 2000]$ Hz is mapped to more than 1500 mel while the interval $[8000, 10000]$ Hz is mapped to less than 300 mel.

\newcommand\mel{2595*log10(1+x/700)} 

\begin{figure}
    \centering
    \begin{tikzpicture}
    
    \begin{axis}[every axis plot post/.append style={
      mark=none,domain=0:10000,samples=200,smooth}, 
    axis x line*=bottom, 
    axis y line*=left, 
    enlargelimits=upper,
    x label style={at={(axis description cs:0.5,-0)},anchor=north},
    y label style={at={(axis description cs:.05,.5)},anchor=south},
    xlabel={Frequency (hertz)},
    ylabel={Mel sclae (mel)}] 
    \addplot {\mel};
    \end{axis}
    \end{tikzpicture}

    \caption{Mel scale representing the perception of frequencies}
    \label{mel_scale}
\end{figure}
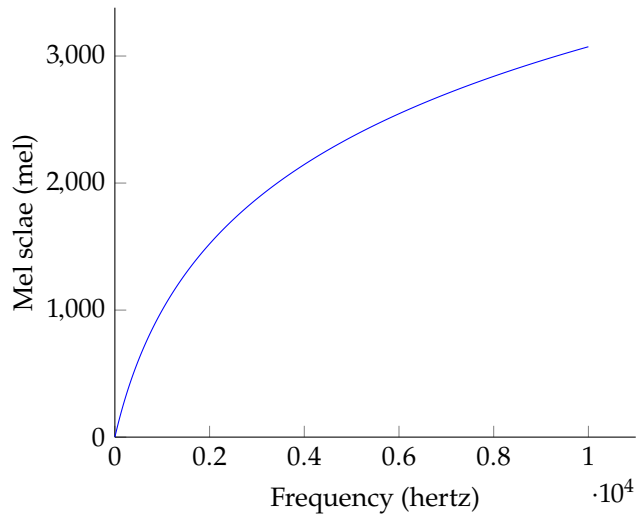




\section{Modeling of Emotion Expressiveness}

Emotion modeling is one of the main challenges in developing more natural human-machine interfaces. Among the many existing approaches, Two of them are widely used in applications.
A first representation, is Ekman's six basic emotion model~\citep{basic_emotions-92-Ekman} which identifies anger, disgust, fear, happiness, sadness and surprise as six basic categories of emotions from which the other emotions may be derived.\\

Emotions can also be represented in a multidimensional continuous space like in the Russels circumplex model~\citep{circumplex-80-russel}. This model makes it possible to better reflect the complexity and the variations in the expressions, unlike the category system. The two most commonly used dimensions in the literature are arousal, corresponding to the level of excitation and valence, corresponding to the pleasantness level or positiveness of the emotion. A third dimension is sometimes added:  dominance corresponding to the level of power of the speaker relative to the listener.\\

A more recent way of representing emotions is based on ranking which prefers a relative preference method to annotate emotions rather than labeling them with absolute values~\cite{yannakakis17ordinalemotions}. The reason is that humans are not reliable for assigning absolute values to subjective concepts. However they are better at discriminating between elements shown to them. Therefore, the design of perception tasks, e.g. about emotion or style in speech, should take this into account by asking participants to solve comparison tasks rather than rating tasks.


It is important to note that many other approaches exist~\cite{theory_constructed_emotions-17-barrel} and it is a difficult question to know what approach should be used in applications in the field of Human-Computer Interaction. Indeed these psychological models of affect are propositions of explanations of how emotions are expressed. But these propositions are difficult to assess in practice.





Humans express their emotions via various channels: face, gesture, speech, etc. %
Different people will express and perceive emotions differently depending on their personality, their culture and many other aspects. For developing application, one has therefore to take assumptions to reduce its scope and choose one approach of emotion modeling.

In this Chapter we are interested in how the expressive speech synthesized will be perceived. It is therefore reasonable to begin by choosing a language and assuming the origin of the synthesized voice.

Research has recently evolved into systems using, without preprocessing, the signal or spectrogram of the signal as input: the neural network learns the features that best correspond to the task it is supposed to perform on its own. This principle has been successfully applied to the modeling of emotions, currently constituting the state of the art in speech emotion recognition~\citep{affect_modeling-13-martinez, adieu_features-16-trigeorgis}.\\

\section{Expressive Speech Synthesis}


\subsection{A brief History of Speech Synthesis Techniques and How to Control Expressiveness}



    

The goal behind a speech synthesis system is to generate an audio speech signal corresponding to any input text. 

A sentence is constituted of characters and a human knows how these characters should be pronounced. If we want a machine to be able to generate speech signal from text, we have to teach it, or program it to do the same.

Such systems have been developed for decades and many different approaches were used. Here we summarize them in three categories: Concatenation, Parametric Speech Synthesis and Statistical Parametric Speech Synthesis. However the state of the art is more diverse and complex. It contains many variants and hybrid approaches between them.

\subsubsection{Concatenation}
This approach is based on the concatenation of pieces of audio signals corresponding to different phonemes.
This method is segmented in several steps. 
First, the characters should be converted in the corresponding phones to be pronounced. A simplistic approach is to assume that one letter corresponds to one phoneme for example. Then the computer must know what signal corresponds to a phoneme. A possibility to solve this problem is to record a database containing all the existing phonemes in a given language.

However concatenating phones one after another leads to very unnatural transitions between them. 
In the literature, this problem was tackled by recording successions of two phonemes, called diphones, instead of phones. All combinations of diphones are recorded in a dataset. The generation of speech is then performed by concatenation of these diphones. 

In this approach, many assumptions are not met in practice.

First, a text processing has to be performed. Indeed, text is constituted of punctuation, numbers, abbreviations, etc.
Moreover, the letter to sound relationship is not respected in English and in many other languages. The pronunciation of words often depend on the context.
Also, concatenating phone leads to a chopped signal and prosody of the generated signal is unnatural. 

To have a control on expressiveness with diphone concatenation techniques, it is possible to change $F0$ and duration with signal processing techniques implying some distorsion on the signal. Other parameters cannot be controlled without altering the signal leading to unnatural speech.

Another approach that is also based on the concatenation of pieces of signal is Unit Selection. Instead of concatenating phones (or diphones), larger parts of words are concatenated. An algorithm has to select the best units according to criteria: few discontinuities in the generated speech signal, a consistent prosody, etc.

For this purpose, a much larger dataset must be recorded containing a large variety of different combinations of phone series. The machine must know what part of signal corresponds to what phoneme, which means it has to be annotated by hand accurately. This annotation process is time consuming. Today there exist tools to do this task automatically. But this automation can in fact be done at the same time as synthesis as we will see later.

The advantages of this method is that the signal is less altered and most of the transitions between phones are natural because they are coming as is from the dataset.

With this method, a possibility to synthesize emotional speech is to record a dataset with separate categories of emotion. In synthesis, only units coming from a category will be used~\cite{emotional_speech_synthesis-01-schroder}. The drawback is that it is limited to discrete categories without any continuous control.

\subsubsection{Parametric Speech Synthesis} 

Parametric Speech Synthesis is based on modeling how the signal is generated. It allows interpretability of the process. But in general, simplistic assumptions have to be made for speech modeling.


Anatomically, the speech signal is generated by an excitation signal generated in the larynx. This excitation signal is transformed by resonance through the vocal
tract which acts as a filter constituted by the guttural, oral and nasal cavities. If this excitation signal is generated by glottal pulses, then a voiced sound is obtained. Glottal pulses are generated by a series of openings and closures of vocal cords or vocal folds. The vibration of the vocal chords has a fundamental frequency.

As opposed to voiced sounds, when the excitation signal is a simple flow of exhaled air, it is an unvoiced sound.

\begin{figure}[!ht]
	\centering
	\includegraphics[width=1\columnwidth]{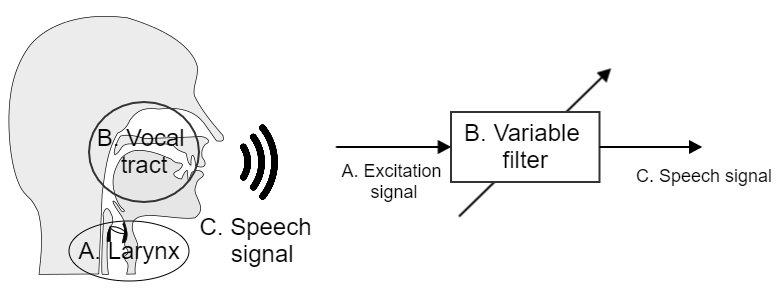}
	\centering
	\caption{Diagram describing voice production mechanism and source-filter model}
	\label{voice_production}
\end{figure}

The source-filter model is a way to represent speech production which uses the idea of separating the excitation and the resonance phenomenon in the vocal tract. It assumes that these two phenomena are completely decoupled. The source corresponds to the glottal excitation and the filter corresponds to the vocal tract. This principle is illustrated in Figure \ref{voice_production}\footnote{Vocal tract image from: \url{https://en.wikipedia.org/wiki/User:Tavin##/media/File:VocalTract.svg} - Tavin/CC-BY-3.0}.

An example of Parametric Speech modeling is the linear prediction model. 
The linear prediction (LP) model uses this theory assuming that the speech is the output signal of a recursive digital filter, when an excitation is received at the input. In other words, it is assumed that each sample can be predicted by a linear combination of the last $p$ samples. The linear predictive coding works by estimating the coefficients of this digital filter representing the vocal tract. The number of coefficients to represent the vocal tract has to be chosen. The more coefficients we take, the better the vocal tract is represented, but the more complex the analysis will be. The excitation signal can then be computed by applying the inverse filter on the speech signal. 

In synthesis, this excitation signal is modeled by a train of impulses. In reality, the mechanics of the vocal folds is more complex making this assumption too simplistic.

The vocal tract is a variable filter. Depending on the shape we give to this vocal tract, we are able to produce different sounds. A filter is considered constant for a short period of time and a different filter has to be computed for each period of time. 

This approach has been successful to synthesize intelligible speech but not natural human sounding speech. 

For expressive speech synthesis, this technique has the advantage of giving access to many parameters of speech allowing a fine control. 

The approach used in~\cite{burkhardt2000verification_acoustical_correlates_emotion} to discover how to control a set of parameters to obtain a desired emotion was done through perception tests. 
A set of sentences were synthesized with different values of these parameters. These sentences were then used in listening tests in which participants were asked to answer questions about the emotion they perceived. Based on these results, values of the different parameters were associated to the emotion expressions.

\subsubsection{Statistical Parametric Speech Synthesis}

Statistical Parametric Speech Synthesis (SPSS) is less based on knowledge, and more based on data. It can be seen as Parametric Speech synthesis in which we take less simplistic assumptions on the speech generation and rely more on the statistics of data to explain how to generate speech from text.


The idea is to teach a machine the probability distributions of signal values depending on the text that is given. We generally assume that generating the values that are most likely is a good choice. We thus use the Maximum Likelihood principle (see Section~\ref{MLE}).

These probability distributions are estimated based on a speech dataset. To be a good estimation of the reality, this dataset must be large enough. 

The first successful SPSS systems were based on hidden Markov models (HMMs) and Gaussian Mixture models (GMMs).


The most recent statistical approach uses DNN~\citep{stat_param_speech_synthesis_dnn-13-zen} which is the basis of new speech synthesis systems such as WaveNet~\citep{wavenet-16-vandenoord} and Tacotron~\citep{tacotron-17-wang}. The improvement provided by this technique~\citep{hmms_to_dnns-16-watts} comes from the replacement of decision trees by DNNs and the replacement of state prediction (HMM) by frame prediction. 

In the rest of this chapter, we focus on this approach of Speech Synthesis. Section~\ref{deep_learning} explains Deep Learning focusing in Speech Synthesis application and Section~\ref{information_theory} 


\subsubsection{Summary}
Depending on the synthesis technique used~\citep{emotional_speech_synthesis-14-Burkhardt}, the voice is more or less natural and the synthesis parameters are more or less numerous. These parameters allow to create variations in the voice. The number of parameters is therefore important for the synthesis of expressive speech.

While parametric speech synthesis can control many parameters, the resulting voice is unnatural. 
Synthesizers using the principle of concatenation of speech segments seem more natural but allow the control of few parameters.

The statistical approaches allow to obtain a natural synthesis as well as a control of many parameters~\citep{statistical_param_speech_syn-09-zen}.

\subsection{Deep Learning for Speech Synthesis}
\label{deep_learning}
Machine Learning consists of teaching a machine to perform specific task, using data. In this Chapter, the task we are interested in is Controllable Expressive Speech Synthesis.

The mathematical tools for this come from the field of Statistical Modeling.



Deep Learning is the optimization of a mathematical model which is a parametric function with many parameters. This model is optimized or \textit{trained} by comparing its predictions to ground truth examples taken from a dataset. This comparison is based on a measure of similarity or error between a prediction and the true example of the dataset. The goal is then to minimize the error or maximize the similarity. This can always be formulated as the minimization of a loss function.

To find a good loss function, it is necessary to understand the statistics of the data we want to predict and how to compare them. For this, concepts from information theory are used.





\subsubsection{Different operations and Architectures}

The form of the mathematical function used to process the signal can be constituted of lots of different operations. Some of these operations were found very performant in different fields and are widely used. In this Section, we describe some operations relevant for speech synthesis. In Deep Learning, the ensemble of the operations applied to a signal to have a prediction is called \textit{Architecture}. There is an important research interest in designing architectures for different tasks and data to process. This research reports empirical results comparing the performance of different combinations. The progress of this field is directly related to the computation power available on the market.

Historically, the root of Deep Learning is a model called Neural Network.  This model was inspired by the role of neurons in brain that communicate with electrical impulses and process information.

Since that, more recent models drove away from this analogy and evolves depending on their actual performance.

Fully connected neural network, are successions of linear projections followed by non-linearities (sigmoid, hyperbolic tangent, etc.) called layers.

\begin{equation}
    h=f_{h}(W_{h}x+b_{h})
\end{equation}

$x$: input vector

$h$: hidden layer vector

$W$ and $b$: parameter matrices and vector

$f _{h}$: Activation functions

\paragraph{}

More layers implies more parameters and thus a more complex model. It also means more intermediate representations and transformation steps. It was shown that deeper Neural Networks (more layers) performed better than shallow ones (fewer layers). This observation lead to the names Deep Neural Networks (DNNs) and Deep Learning. A complex model is capable of modeling a complex task but is also more costly to optimize in terms of computation power and data.

Merlin~\cite{merlin-16-wu} toolkit has been an important tool to investigate the use of DNNs for speech synthesis. The first models developed within Merlin were based only on Fully connected neural networks. One DNN was used to predict acoustic parameters and another one to predict phone durations. It was a first successful attempt that outperformed other statistical approaches at the time.

Time dependencies are not well modeled and it ignores the autoregressive nature of speech signal. In reality, this approach relies a lot on data and do not use enough knowledge. 


Convolutional Neural Networks (CNNs) refer to the operation of convolution and remind the convolution filters of signal processing (see Equation~\ref{eq_conv_filter}). A convolution layer can thus be seen as a convolutional filter for which the coefficients were obtained by training the Deep Learning architecture.

\begin{equation}
\label{eq_conv_filter}
    g(x,y)=\omega *f(x,y)=\sum _{s=-a}^{a}{\sum _{t=-b}^{b}{\omega (s,t)f(x-s,y-t)}}
\end{equation}

$f$: input matrix

$g$: output matrix

$\omega$: convolutional filter weights

\paragraph{}

Convolutional filters were studied in the field of image processing. We know what filters to apply to detect edges, to blurr an image, etc.

In practice, often, the operation implemented is correlation which is the same operation except that the filter is not flipped. Given that the parameters of the filters are optimized during training, the flipping part is useless. We can just consider that the filter optimized with a correlation implementation is just the flipped version of the one that would have been computed if convolution was implemented.

For speech synthesis, convolutional layers have been used to extract a representation of linguistic features and predict spectral speech features.

For a temporal signals such as speech, one dimensional convolution along the time axis allows to model time dependencies. As layers are stacked, the receptive field increase proportionally. In speech, there are long-term dependencies in the signal, e.g. in the intonation and emphasis of some words. To model these long-term dependencies, dilated convolution was proposed. It allows to increase the receptive field exponentially instead of proportionally with the number of layers.


Recurrent Neural Network involves a recursive behaviour, i.e. having an information feedback from the output to the input. This is analogous to recursive filters. Recursive filters are filters designed for temporal signals because they are able to model causal dependencies. It means that at a give time $t$, the value depends on the past values of the signal.

\begin{equation}
h_{t}=f_{h}(W_{h}x_{t}+U_{h}h_{t-1}+b_{h})
\end{equation}

\begin{equation}
    y_{t}=f_{y}(W_{y}h_{t}+b_{y})
\end{equation}

$x_{t}$: input vector

$h_{t}$: hidden layer vector

$y_{t}$: output vector

$W$, $U$ and $b$: parameter matrices and vector

$f _{h}$ and   $f_y$: Activation functions

\subsubsection{Encoder and Decoder}

An encoder is a part of neural network that outputs a hidden representation (or latent representation) from an input. A decoder is a part of neural network that retrieves an output from a latent representation.

When the input and the output is the same, we talk about auto-encoders. The task in itself is useless, but the interesting part here is the latent representation. The latent space of an auto-encoder can provide interesting properties such as a lower dimensionality, meaning a compressed representation of the initial data or meaningful distances between examples.

\subsubsection{Sequence-to-sequence modeling and Attention Mechanism}

A sequence-to-sequence task is about converting sequential data from one domain to another, e.g. from a language to another (translation), from speech to text (speech recognition) or from text to speech (speech synthesis).

First Deep Learning architectures for solving sequence-to-sequence tasks were based on encoder-decoder with RNNs called RNN transducer.

Other techniques were were found to outperform this. The use of Attention Mechanism was found beneficial~\cite{seq2seq_comparison-17-Prabhavalkar}.

Attention Mechanism was first developed in the field of computer vision. It was then successfully applied to Automatic Speech Recognition (ASR) and then to Text-to-Speech synthesis (TTS).

In the Deep Learning architecture, a matrix is computed and used as weighting on the hidden representation at a given layer. The weighted representation is fed to the rest of the architecture until the end. This means that the matrix is asked to emphasize the part of the signal that is important to reduce the loss. This matrix is called the Attention matrix because it represents the importance of the different regions of the data.

In computer vision, a good illustration of this mechanism is that for a task of classification of objects, the attention matrix has high weights for the region corresponding to the object and low weights corresponding to the background of the image.

In ASR, this mechanism has been used in a so called  \textit{Listen, Attend and Spell}~\cite{listen_attend_spell-16-chan} (LAS) setup. An important difference compared to the previous case that it is a sequential problem. There must be an information feedback to have a recursive kind of architecture and each time step must be computed based on previous time steps.

LAS designate three parts of the Deep Learning architecture. The first one encodes audio features in a hidden representation. The role of the last one is to generate text information from a hidden representation. Between this encoder and decoder, at each time step, an Attention Mechanism computes a vector that will weight the text encoding vector. This weighting vector should give importance to the part of the utterance that the architecture should pay attention at to generate the corresponding part of speech.

\begin{figure}[!ht]
	\centering
	\includegraphics[width=1\columnwidth]{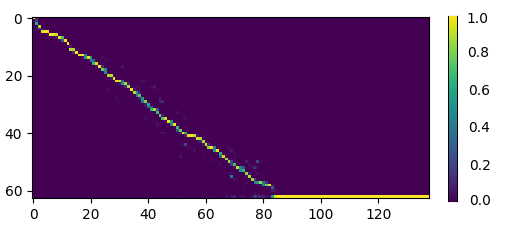}
	\centering
	\caption{Alignment plot. The y-axis represents the character indices and the x-axis represents the audio frame feature indices. The color scale corresponds to the weight given to a given character to predict a given audio frame.}
	\label{alignment}
\end{figure}

An Attention plot (see Figure~\ref{alignment}) of a generated sentence can be constructed by juxtaposing all the weighting vectors computed during the generation of a sentence. The resulting matrix can then be represented by mapping a color scale on the values contained. 

This attention plot shows an attention path, i.e. the importance given to characters along the audio output timeline. As it can be observed in Figure~\ref{alignment}, this attention path should have a close to diagonal shape. Indeed the two sequences have a close chronological relationship.

\subsection{Information Theory and Speech Probability distributions}
\label{information_theory}

Information Theory is about optimizing how to send messages with as few resources as possible. To that end, the goal is to compress the information by using the right code so that the messages do not contain redundancies to be as small as possible.

\subsubsection{Information and probabilities}

Shannon's Information Theory quantifies information thanks to the probability of outcomes. If we know an event will occur, its occurrence gives no information. The less likely it is to happen, the more it gives information.

This relationship between information and probability of an event is given by Shannon information content measured in bits. A bit is a variable that can have two different values: 0 or 1.
\begin{equation}
    h(x)=\log_2\left(\frac{1}{p(x)}\right)
    \label{information_content}
\end{equation}

The number of possible messages with $L$ bits is $2^L$. If all messages are equally probable, the probability of each message is $p=\frac{1}{2^L}$. We then have $L=\log_2\left(\frac{1}{p}\right)$. A generalization of this formula in which the messages are not equally probable is Equation~\ref{information_content}. It can be interpreted as the minimal number of bits to communicate this message.

The probability represents the degree of belief that an event will happen~\cite{information_theory_inference-03-mackay}. For example, we can wonder the probability of a result of four by rolling a six sided die or the probability of that the next letter in a text \textit{r}.


These probabilities depend on the assumptions we make: 
\begin{itemize}
    \item Is the die perfectly balanced? If yes, the probability of a result of four is $1/6$. 
    \item What is the language of the text ? Do we know the subject, etc. Depending on these information we can have different estimations of this probability.
\end{itemize}

We obtain a probability distribution by listing the probability of all the possible outcomes. For the example of the result by rolling the perfectly balanced die, the possible outcomes are $[1,2,3,4,5,6]$ and their probabilities are $[1/6,1/6,1/6,1/6,1/6,1/6]$.

In both examples, we have a finite number of possible outcomes. The probability distribution is said discrete. On the contrary, when the possible outcomes are distributed on a continuous interval, then the probability distribution is said continuous. This is the case, for example, of amplitude values in a spectrogram.


The most famous continuous probability distribution is the Gaussian distribution:
\begin{equation}
    p(x)={\frac {1}{\sigma {\sqrt {2\pi }}}}\operatorname {e} ^{-{\frac {1}{2}}\left({\frac {x-\mu }{\sigma }}\right)^{2}}
\end{equation}

Another important distribution, especially in speech processing is Laplacian distribution:

\begin{equation}
    p(x)={\frac {1}{2b}}\operatorname{e} ^ {\left(-{\frac {|x-\mu |}{b}}\right)}\,\!
\end{equation}

\newcommand\gauss[2]{1/(#2*sqrt(2*pi))*exp(-((x-#1)^2)/(2*#2^2))} 
\newcommand\laplace[2]{1/(#2*2)*exp(-(abs((x-#1)))/(#2))} 

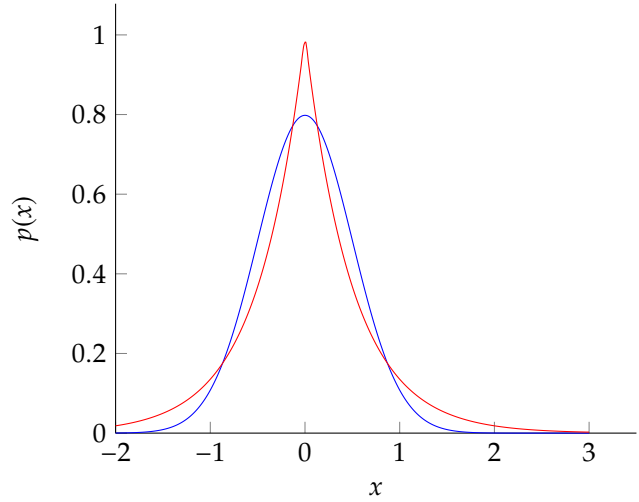
\begin{figure}
    \centering
    \begin{tikzpicture}
    \begin{axis}[every axis plot post/.append style={
      mark=none,domain=-2:3,samples=200,smooth}, 
    axis x line*=bottom, 
    axis y line*=left, 
    enlargelimits=upper,
    x label style={at={(axis description cs:0.5,-0)},anchor=north},
    y label style={at={(axis description cs:.05,.5)},anchor=south},
    xlabel={$x$},
    ylabel={$p(x)$}] 
    \addplot {\gauss{0}{0.5}};
    \addplot {\laplace{0}{0.5}};
    \end{axis}
    \end{tikzpicture}

    \caption{In blue: Gaussian distribution with $\mu=0$ and $\sigma=0.5$. In Red: Laplacian distribution with $\mu=0$ and $b=0.5$}
    \label{distributions}
\end{figure}

Both distributions are plotted in Figure~\ref{distributions}. The blue curve corresponds to the Gaussian probability distribution (with $\mu=0$ and $\sigma = 0.5$) and the red curve corresponds to the Laplacian probability distribution (with $\mu=0$ and $b = 0.5$). For both distributions, the maximum is $\mu$ are symmetrically decreasing as the distance from $\mu$ increases.

\subsubsection{Entropy and relative-entropy}
The average information content of an outcome, also called entropy, of the probability distribution $p$ is:

\begin{equation}
    H(p) = \sum_x p(x)\log_2\left(\frac{1}{p(x)}\right)
\end{equation}

The relative-entropy between two probability distributions, also called Kullback-Leibler divergence is defined as:
\begin{equation}
    D_{\mathrm {KL} }(p\|q)=\sum_x p(x)\log {\frac {p(x)}{q(x)}}\;dx
\end{equation}

It represents a dissimilarity between two probability distributions.



\subsubsection{Maximum likelihood and particular cases}
\label{MLE}

This concept is necessary to understand how to train a Deep Learning algorithm or more generally, how to find the optimal parameters of a model. The role of a statistical model is to represent as accurately as possible the behaviour of a probability distribution.

Maximum likelihood estimation (MLE) (see Equation~\ref{eq_MLE}) allows to estimate the parameters $\theta$ of a statistical parametric model $p(x|\theta)$ by maximizing the probability of a dataset under the assumed statistical model, i.e. the Deep Learning architecture.

\begin{equation}
\label{eq_MLE}
    \theta_{MLE} = \arg\max p(\mathbf{x}|\theta)
\end{equation}

It can be demonstrated this is equivalent to minimizing $D_{\mathrm {KL} }(p\|q)$ with $p$, the probability distribution of the model and $q$, the probability of the data~\cite{deep_learning-16-goodfellow}. It is a way to express that the probability distribution generated by the model should be as close as possible to the probability distribution of the data.

If assumptions can be made on the probability distributions, it is possible to have distances or errors for which the minimization is equivalent to MLE. These errors are computed by comparing estimations from the model ${\hat {Y_{i}}}$ and the value from the dataset $Y_{i}$.


Maximizing likelihood assuming a Gaussian distribution is equivalent to minimizing Mean Squared Error (MSE):

\begin{equation}
    {MSE} ={\frac {1}{n}}\sum _{i=1}^{n}(Y_{i}-{\hat {Y_{i}}})^{2}
\end{equation}



Maximizing likelihood assuming a Laplacian distribution is equivalent to minimizing Mean Absolute Error (MAE):

\begin{equation}
    {MAE} ={\frac {1}{n}}\sum _{i=1}^{n}\left|Y_{i}-{\hat {Y_{i}}}\right|
\end{equation}


To choose the right criterium to optimize when working with speech data, one should pay attention to speech probability distributions.
Speech waveforms and magnitude spectrogram distribution are Laplacian~\cite{speech_probability_distribution-03-gazor, probabilistic_modeling_speech-18-usman}. 
That is why MAE loss should be used to optimize their predictions.






\section{Summary and Application}



In this Chapter, we first briefly introduced digital signal processing and digital filtering, and described the different possibilities of emotion representation and the few most important speech feature spaces in this context, namely spectrogram and mel-spectrogram.

Available speech synthesis methods were then  exposed, from concatenation of speech signal segments to parametric modelling of speech production, to statistical parametric speech synthesis.

Most recent SPSS systems use Deep Learning that can be seen as non-linear signal processing for which filters as optimized based on data.

We focused on the tools for SPSS and explained Deep Learning architecture blocks that are used along with the right loss functions based on the probability distributions of speech features.

To build a controllable expressive speech synthesis system, one should keep several concepts in mind. First, it is necessary to gather data and process them to have a good representation to be used with a Deep Learning algorithm, i.e. text, mel-spectrograms, and information about the expressivness of speech. Then one has to design a Deep Learning architecture. Its operations should be inspired by the features to model (1D convolution or RNN cells for long term context, attention mechanism for recursive relationships). It should have a way to control expressiveness either with a categorical representation~\cite{exploring_transfer_learning-19-tits} or a continuous representation~\cite{visualization-19-tits}. But it is important to take into account that annotations should not be acquired from humans by asking them to give absolute values on subjective concepts, but rather by asking them to compare examples. And finally, the parametric model should be trained with a loss function adapted to the probability distribution of the acoustic features, i.e. MAE and Kullback-Leibler divergence loss.

\begin{backmatter}

\section*{Acknowledgments}
No\'e Tits  is funded through a PhD grant from the Fonds pour la Formation \`a la Recherche dans l'Industrie et l'Agriculture (FRIA), Belgium.

\begin{authordetails}
	
	
	\author{No\'e Tits$^{1,*}$, Kevin El Haddad$^{1}$ and Thierry Dutoit$^1$}
	\address[1]{Numediart Institute, University of Mons, Mons, Belgium}
	%
	\address{*Address all correspondence to: noe.tits@umons.ac.be}
	%
	
	\IntechOpentext{\textcopyright\ \the\year{} The Author(s). License IntechOpen. This chapter is distributed under the terms of the Creative Commons Attribution License (http://creativecommons. org/licenses/by/3.0), which permits unrestricted use, distribution, and reproduction in any medium, provided the original work is properly cited.}
	
	
\end{authordetails}

\bibliographystyle{vancouver}
\bibliography{refs}

\end{backmatter}

\end{document}